\documentclass[12pt,a4paper]{article}

\usepackage{amscd,amsfonts,amsmath,amssymb,amsthm,bbm,bm,latexsym,mathrsfs}
\usepackage{epsfig,graphics,natbib,graphicx,subfigure,longtable}%, algorithm2e}
\usepackage[english]{babel}
\usepackage[usenames]{color}
\usepackage{booktabs}
\usepackage{sgame, float}
\usepackage[top=1.7in, bottom=1.7in, left=0.9in, right=0.9in]{geometry}
\usepackage{epstopdf}
\usepackage{algorithmicx}
\usepackage{algpseudocode}
\usepackage{algorithm}
\usepackage{rotating}

\allowdisplaybreaks
\makeatother

\newfloat{algorithm}{tbp}{loa}
\providecommand{\algorithmname}{Algorithm}
\floatname{algorithm}{\protect\algorithmname}

\begin{document}

\title{\vspace{-50pt} Modelling Preference Data with the Wallenius Distribution}

\author{
\hspace{-20pt} 
Clara Grazian\textsuperscript{1} \hspace{15pt}
Fabrizio Leisen\textsuperscript{2} \hspace{15pt}
Brunero Liseo\textsuperscript{3}
        \\ 
        \vspace{5pt}
        \\
        {\centering {\small
        \textsuperscript{1}
        University of Oxford, U.K. \hspace{18pt}  
         \textsuperscript{2} 
        University of Kent, U.K. 
        }} \\
        {\centering {\small
        \textsuperscript{3}
         Sapienza Universit\`a di Roma, Italy \hspace{18pt}
        }} \vspace{5pt} \\
     }

\date{}
\maketitle

%%%%%%%%%%%%%%%%%%%%%%%%%%%%%%%%%%%%%%%%%%%%%%%%%%%%%%
%%         	Abstract			%%%
%%%%%%%%%%%%%%%%%%%%%%%%%%%%%%%%%%%%%%%%%%%%%%%%%%%%%%
\abstract{ %Ranking datasets is useful when statements on the order of observations are more important than the magnitude of their differences and little is known about the underlying distribution of the data.
The Wallenius distribution is a generalisation of the Hypergeometric
distribution where weights are assigned to balls of different colours.
This naturally defines a model for ranking categories which can be
used for classification purposes. Since, in general,
the resulting likelihood is not analytically available, we adopt an approximate Bayesian computational (ABC) approach 
for estimating the importance of the categories. We illustrate the performance of the estimation procedure on simulated datasets. Finally, we use the new model for analysing two datasets concerning movies ratings and Italian academic statisticians' journal preferences. The latter
is a novel dataset collected by the authors.

\textbf{Keywords}: Approximate Bayesian Computation, Biased Urn, Movies ratings, Scientific Journals Preferences. 
}

%%%%%%%%%%%%%%%%%%%%%%%%%%%%%%%%%%%%%%%%%%%%%%%%%%%%%%
%%    	     	Intro			%%%
%%%%%%%%%%%%%%%%%%%%%%%%%%%%%%%%%%%%%%%%%%%%%%%%%%%%%%
\section{Introduction and motivations}
\label{Intro}

Human beings naturally tend, in everyday life, to compare and rank concepts and objects such as food, shops, singers and football teams, according to their preferences. In general, to rank a set of objects means to arrange them in order with respect to some characteristic. Ranked data are often employed in contexts where objective and precise measurements are difficult, unreliable, or even impossible to obtain and the observer is bound to collect ordinal information about preferences, judgments, relative or absolute ranking among competitors, called items. 
Modern web technologies have made available a huge amount of ranked data, which can provide information about social and psychological behaviour, marketing strategies and political preferences. The codification of this information has been of interest to the statisticians since the beginning of the 20th century. The \textit{Thurstone model} (TM) assumes that each
item $i$ is associated with a score $W_i$ on which the comparative judgment is based; examples of unidimensional scores are the unrecorded finishing times of players in a race or any possible preference/attitude measure towards items.
Item $i$ is preferred to item $j$ if $W_i$ is greater than $W_j$, see \cite{thurs}. From the modelling point of view, this corresponds to assigning a probability $p_{ij}=\Pr(W_i > W_j)$. 
The \textit{Bradley-Terry} model (BT) is a particular case of the TM model with $p_{ij}=p_i(p_i+p_j)^{-1}$ where $p_i,p_j\geq 0$  are the item parameters reflecting the rate of each item, see \cite{bt52}. Paired comparison models are always applicable to rankings after converting the latter in a suitable set of pairwise preferences. Conversely, paired comparisons of K items do not necessarily correspond to a ranking, due to the potential presence of circularities. A popular extension of the BT model is the \textit{Plackett-Luce model} (PL). 
Given a set of $L$ items and a vector of probabilities $(p_1,\dots, p_L)$, such that $\sum_{i=1}^L p_i=1$, the PL model assigns a probability distribution on all the set of possible rankings of these objects which is a function of the $(p_1,\dots, p_K)$, see \cite{plack} and \cite{luce}. TM, BT and PL are not the only proposals in the field, and modelling ranking is an active area of research, see \cite{marden} and \cite{alvo}.

There is no wide consensus about the use of choice or ranking data for better representing preferences and, very often, the best solution is problem specific. In this paper, we consider a sort of hybrid situation; in fact, we assume that choices related to single items can be further classified into categories of different relevance, and the ranking of categories is the main goal of the statistical analysis. 
Our approach makes use of an extension of the Hypergeometric distribution, namely the Wallenius distribution \citep{walle} and can be used in the cases where data are available in the form of rankings, votes, preferences of items but the interest is in defining the importance of the categories in which the items can be 
clustered.

The Wallenius distribution arises quite naturally in situations where sampling is performed without replacement and units in the population have different probabilities to be drawn. To be more specific, consider a urn with balls of $c$ different colours: for $i = 1, \dots, c$ there are $m_i$ balls of colour $i$. In addition, colour $i$ has a priority $\omega_i>0$ which specifies its relative importance with respect to the other colours. A sample of $n$ balls, with $n < \sum_{i=1}^c m_i$, is drawn sequentially without replacement. The Wallenius distribution describes the probability distribution for all possible strings of balls of length $n$ drawn from this urn. 
This experimental situation arises in very different contexts. For example, in auditing problems, transactions are examined by randomly selecting a single euro (or pound, or dollar) among the total amount, so larger transactions are more likely to be drawn and checked. 

The Wallenius distribution was introduced by \cite{walle} and it is also known as the noncentral Hypergeometric distribution; this alternative name is justified by the fact that, when all the priorities $\omega_i$'s are equal, one gets back to the classical Hypergeometric distribution. However this name should be avoided because, as extensively discussed by \cite{fog1}, this is also the name of another distribution, proposed by \cite{fisher1}.
Although the Wallenius distribution is a very natural statistical model for the aforementioned situations, its popularity in applied settings has been prevented by the lack of a closed form expression of the probability mass function: see Section \ref{Prelim} for details.

The gist of this paper is the use of the priorities vector $\bm{\omega}=(\omega_1, \dots, \omega_c)$ of the Wallenius distribution as a measure of importance for different values of a categorical variable. 

In particular, we analyse two datasets, where we aim at ranking the categories rather than the items.
The first dataset considers data downloaded from the MovieLens website, which consists of 105,339 ratings across 10,329 movies performed by 668 users.
In this framework, it is of interest to classify the different genres in terms of  satisfaction, in order to provide some useful feedback to users and/or providers.

The second dataset considers data we collected between October and November 2016 among Italian academic statisticians.
They indicated their journal preferences from the 2015 ISI ``Statistics and Probability'' list of Journals. In this context, we are interested in ranking the journal categories in order to provide a description of the research interests of the Italian Statistical community.

We adopt a Bayesian methodology which allows us to overcome the computational problems related to the lack of a closed form expression of the probability mass function of the Wallenius distribution.
We propose a novel approximate Bayesian computational approach \citep{marin-abc}, where the vector of summary statistics is represented by the relative frequencies of the different categories and the acceptance mechanism is based on the distance in variation \citep{bremaud}

The paper is organized as follows: in Section \ref{Prelim} we introduce the Wallenius distribution; in Section \ref{BayesInf} our approximated inferential strategy is described, based on an ABC algorithm. The performance of the algorithm has been tested in several examples, first in an extensive simulation study (Section \ref{Simu}) and then on two real datasets (Section \ref{RealData}). A discussion concludes the paper. 

%%%%%%%%%%%%%%%%%%%%%%%%%%%%%%%%%%%%%%%%%%%%%%%%%%%%%%
%% 	 	Preliminaries		%%%
%%%%%%%%%%%%%%%%%%%%%%%%%%%%%%%%%%%%%%%%%%%%%%%%%%%%%%
\section{The Wallenius Distribution}
\label{Prelim}

Consider an urn with $N$ balls of $c$ different colours.  There are $m_i$ balls of the $i$-th colour, so that $\sum_i^c m_i=N$. In this situation, the multivariate Hypergeometric distribution is the discrete probability distribution which describes the sampling without replacement of $n$ balls. In this framework, the probability of drawing a ball of a certain colour is proportional to the number of balls of the same colour. It is possible to generalise the experiment with a biased sampling of balls. For instance, each colour may have a different priority or importance, say $\omega_i>0$, $i=1, \dots, c$. Suppose we have drawn $n$ balls without replacement from the urn and let $\bm{X}_n=\left (X_{1n}, X_{2n}, \dots, X_{cn}\right)$ denote the frequencies of balls of different colours in the sample. Let  $Z_n$ be the colour of the ball drawn at time $n$. 
In this setting, the probability that the next ball is of colour $i$ also depends on its priority and is defined as

\begin{equation}
\label{wal-def}
P\left( Z_{n+1} = i \vert \bm{X}_n \right ) = \frac{\left (m_i-X_{in}\right) \omega_{i}}
{\sum_{j=1}^c \left( m_j - X_{jn}\right) \omega_{j} }.
\end{equation}  
\cite{walle} provided the above expression and the probability mass function of $\bm{X}_n$ for the case $c=2$. \cite{chesson}
derived the following general expression. For a given integer $n$, and parameters
$\bm{m}= (m_1, \dots, m_c)$ and $\bm{\omega}= (\omega_1, \dots, \omega_c),$
the probability of observing a vector of colour frequencies $\bm{x}=(x_{1}, \dots , x_{c})$ is 
\begin{equation}
\label{wal-gen}
P(\bm{x}; n, \bm{m, \omega}) = \prod_{j=1}^c \binom{m_j}{x_j} \int_0^1 
\prod_{j=1}^c \left ( 1- t^{\omega_j/d} \right )^{x_j} dt,
\end{equation}
where $\sum_{i=1}^c x_i=n$ and $d= \sum_{j=1}^c \omega_j (m_j - x_j).$ When $\omega_i = \omega$, for every $i=1,\dots, c$, the Wallenius distribution reduces to the multivariate Hypergeometric distribution. This can be easily shown by considering, without loss of generality, $\omega=1$ and $c=2$. In this particular case, the probability mass function simplifies to  
$$P(x; n, m )= \binom{m}{x}\binom{N-m}{n-x}\int_0^1 \left ( 1- t^{1/d}  \right )^{n}dt.$$
The change of variable $z=t^{1/d}$ leads to
\begin{align*}
P(x; n, m )&= \binom{m}{x}\binom{N-m}{n-x}
d \int_0^1 \left ( 1- z \right )^{n}   z^{d-1}  dz \\ &=\binom{m}{x}\binom{N-m}{n-x} \frac{\Gamma(d+1) \Gamma(n+1)}{\Gamma(n+d+1)}.
\end{align*}
Since $d=N-n$, the probability mass function reduces to
$$P(x; n, m )= {\binom{m}{x}\binom{N-m}{n-x}}\bigg /{\binom{N}{n}},$$
which is the probability mass function of the Hypergeometric distribution when two colours are considered. \\

The Wallenius distribution has been underemployed in the statistical literature mainly 
because the integral appearing in \eqref{wal-gen} cannot be solved in a closed form and numerical approximations are necessary. \cite{fog1} has made a substantial contributions in this direction, providing approximations based either on asymptotic expansions or numerical integration. To our knowledge, the Wallenius distribution has only been used in a limited number of applications, mainly devoted to auditing problems \citep{gillett}, ecology \citep{manly}, vaccine efficacy \citep{castillo2000urn} and modeling of RNA sequences \citep{gao11}. 
In this work, we propose a novel look at the Wallenius distribution and we use it as statistical model, with the goal of ranking the values of a categorical random variable, based on preference data.
This is motivated by the sampling nature of the Wallenius distribution where an importance $\omega_j$ is associated with category $j$. The highest $\omega_j$'s represent the most popular categories. This naturally defines a new model which allows us to rank preferences.

Notice that we are implicitly assuming that all balls of the same colour have the same importance; this may not be the case in some applications: we will discuss this aspect in the final section.

 Recently, the development of social networks and the competitive pressure to provide customized services has motivated many new ranking problems involving hundreds or thousands of objects. 
Recommendations on products such as movies, books and songs are typical examples in which the number of objects is extraordinarily large. In recent years, many researchers in statistics and computer science have developed models to handle such big data. For instance, in Section \ref{RealData} we consider the problem of ranking customer movie choices in terms of genres  such as Comedy, Drama and Science Fiction. We consider data downloaded from the MovieLens website \texttt{(www.grouplens.org)} which consists of 105,339 online ratings of 10,329 movies by 668 raters on a scale of 1-5. We rank the categories by estimating the priority parameters of the Wallenius distribution by using an approximate Bayesian approach.
In particular, in the next section, we introduce a simple ABC algorithm which allows us to avoid the direct computation of the integral in equation \eqref{wal-gen}. 

%%%%%%%%%%%%%%%%%%%%%%%%%%%%%%%%%%%%%%%%%%%%%%%%%%%%%%
%% 	 				%%%
%%%%%%%%%%%%%%%%%%%%%%%%%%%%%%%%%%%%%%%%%%%%%%%%%%%%%%
\section{Bayesian Inference for the Wallenius model}
\label{BayesInf}

Let $\bm{x}_h=(x_{h1}, \dots, x_{hc})$ be a draw of $n_h$ balls from the Wallenius urn described in equation \eqref{wal-gen}, 
where $h=1, \dots, k$ and $\sum_{j=1}^c x_{hj}=n_h$.
In this paper we adopt a Bayesian approach, where the parameter vector $\bm{\omega}$
is considered random. For a given prior distribution $\pi(\bm{\omega})$, the resulting posterior 
is
\begin{equation}
\label{posterior}
\pi(\bm{\omega} \vert \bm{x}_1, \dots, \bm{x}_k ) \propto \pi(\bm{\omega})
\prod_{h=1}^k 
\left [ \int_0^1 \prod_{j=1}^c \left ( 1- t_h^{\omega_j/d_h} \right )^{x_{hj}} dt_h
\right ],
\end{equation}

\noindent with $d_h = \sum_{j=1}^c \omega_j (m_j - x_{hj}).$
Here $k$ represents the sample size, that is, the number of different and conditionally
independent preference lists provided by the interviewees, while $n_h (h=1, \dots, k)$ is the number of items 
selected by the $h$-th interviewee. 
The above posterior distribution depends on 
$k$ different integrals which cannot be reduced to a closed form. 
This makes the implementation of standard Markov Chain Monte\,Carlo (MCMC) methods for estimating $\bm{\omega}$ rather complex.
Indeed, most MCMC methods rely on the direct evaluation of the unnormalized posterior distribution \eqref{posterior}.
Although there are many available routines, in different software packages, to evaluate univariate integrals, we noticed that they lack accuracy especially for large values 
of the $n_h$'s and $\bm{m}$.
We believe that this problem has had a strong negative impact   
on the popularization of the Wallenius distribution despite a need for 
interpretable models in the applied setting. For instance, the Wallenius distribution arises naturally in genetics as an alternative to the Fisher exact test, see \cite{gao11} and the references therein. 

In this section, we propose an algorithm which allows to sample from the posterior distribution introduced in \eqref{posterior}. The algorithm belongs to the class of approximate Bayesian computational (ABC) methods. This approach is philosophically different from the standard MCMC methods since the implementation only requires to draw samples from the generating model for a given parameter value. In the case of the Wallenius distribution, the task of generating draws is not hard, making the use of ABC particularly straightforward. \cite{fog2} provided methods and algorithms to sample from the Wallenius distribution. He also made available a reliable \textsc{R} package, called \texttt{BiasedUrn}, which has been used extensively in this work.  

The ABC methodology can be considered as a (class of) popular algorithms that achieves posterior simulation by avoiding the computation of the likelihood function: see \cite{bea:10}, \cite{marin-abc} and \cite{KL2018} for recent surveys. As remarked by \cite{marin-abc}, the first genuine ABC algorithm was
introduced by \cite{Pritchard} in a population genetics setting.
Explicitly, we consider a parametric model $\{f(\cdot \mid \theta), \theta \in \Theta \}$ and suppose that a dataset
$\bm{y}\in\mathcal{D}\subset\mathbb{R}^n$ is observed. Let
$\varepsilon>0$ be a tolerance level, $\eta$ a summary statistic
(which is often not sufficient) defined on $\mathcal{D}$ and $\rho$ a
distance or metric acting on the $\eta$ space. %(\mathcal{D})$. 
Let $\pi$ be a prior distribution for $\theta$; the ABC algorithm is described in Algorithm 1.

\begin{algorithm}[!h]
\caption{ABC Rejection algorithm}
\label{ABC-rej}

		\begin{algorithmic}[1]
			\For {$l=1,\cdots,T$}
			\Repeat
				\State Generate $\theta'$ from the prior 
				distribution $\pi(\cdot)$
				\State Generate $z$ from the likelihood 
				$f(\cdot \mid \theta')$
			\Until $\rho(\eta(\mathbf{z}),\eta(\mathbf{y}))<\varepsilon$
			\State Set $\theta_l = \theta'$
			\EndFor
		\end{algorithmic}
\end{algorithm}

The basic idea behind the ABC is that, for a small (enough)
$\varepsilon$ and a representative summary statistic, we can obtain a
reasonable approximation of the posterior distribution. The practical implementation of an ABC algorithm requires the selection of a suitable summary statistic, a distance and a tolerance level. In our specific case we summarized the data by using the arithmetic mean of the 
observed and simulated frequency vectors, i.e., at the $\ell$-th iteration of pseudo data generation, we have  

\begin{equation}
\label{sum-stat}
\eta(\bm{x}^{(\ell)}) =\widehat{\bm{p}}^{(\ell)} = \frac 1k \sum_{h=1}^k \bm{p}^{(\ell)}_h, 
\end{equation} 
with 
$$
\bm{p}^{(\ell)}_h = \left (\frac{x_{h1}^{(\ell)}}{n_h}, \dots, \frac{x_{hc}^{(\ell)}}{n_h}\right )
$$
to be compared with the relative frequencies observed in the sample
$$
\eta(\bm{x}^{(t)})=\widehat{\bm{p}}^{(t)}= \frac 1k \sum_{h=1}^k \bm{p}^{(t)}_h.
$$
with
$$
\bm{p}^{(t)}_h =\left ( \frac{x_{h1}}{n_h}, \dots, \frac{x_{hc}}{n_h}\right ).
$$
Since the frequencies $\widehat{\bm{p}}^{(\ell)}=\left( \widehat{p}_{1}^{(\ell)},\dots, \widehat{p}_{c}^{(\ell)}\right)$ and $\widehat{\bm{p}}^{(t)}= \left( \widehat{p}_{1},\dots, \widehat{p}_{c}\right)$ can be interpreted as discrete probability distributions, it is natural to compare them through the ``distance in variation'' \citep{bremaud} metrics
\begin{equation}
\label{eq:dist}
\rho(\widehat{\bm{p}}^{(\ell)},\widehat{\bm{p}}^{(t)})=\frac{1}{2}\sum_{j=1}^c \left|\widehat{p}_{j}^{(\ell)}-\widehat{{p}}_{j}\right|
\end{equation}
Regarding the setting of tolerance level we refer to the Section \ref{Simu} where the algorithm will be tested on simulated data.

\subsubsection*{The prior distribution}

The vector of parameters $\bm{\omega}= (\omega_1, \dots, \omega_c)$ assumes values in $\mathbb{R}_+^c$ and different priors can be considered. However, one must take into account that the priority parameters $\omega_j$ must be interpreted in a relative way. 
In fact, the quantity $d$ in the  p.m.f. of the  Wallenius distribution (defined in equation \eqref{wal-gen}) depends on the priority parameters $\boldsymbol{\omega}$. In particular,
$$d= \sum_{j=1}^c \omega_j (m_j - x_j).$$
If we consider two different vectors $\boldsymbol{\omega}\prime$ and $\boldsymbol{\omega}$ such that $\boldsymbol{\omega}\prime = \kappa \boldsymbol{\omega}$ for $\kappa > 0$, we have that 
\begin{equation}\label{ident}
\frac{\omega_j'}{d'}=\frac{\kappa\omega_j}{\sum_{j=1}^c \kappa \omega_j (m_j - x_j)}=\frac{\omega_j}{\sum_{j=1}^c \omega_j (m_j - x_j)}=\frac{\omega_j}{d}
\end{equation}
where $d'$ and $d$ are computed respectively with $\boldsymbol{\omega}\prime$ and $\boldsymbol{\omega}$. Equation \eqref{ident} implies that the p.m.f. of the Wallenius distribution does not change if we consider the vector of priorities $\boldsymbol{\omega}\prime$ instead of $\boldsymbol{\omega}$. This induces an identifiability issue, which can be resolved by a normalization step. 
From this perspective, the most natural way to follow is to assume that $\sum_{j=1}^{c} \omega_j=1$, and to assume a Dirichlet prior on the normalized vector. Hereafter we will assume that the Dirichlet prior we adopt in the simulations and the real data examples are symmetric (i.e., all the hyperparameters are equal). Our default choice will be to set them all equal to 1, making the prior uniform on its support.
An alternative default choice, especially useful when $c$ is large, is given by $\alpha = 1/c$, as explained in \cite{berger2015}.

\subsubsection*{Alternative computational approaches}

The \textsc{R} package \texttt{BiasedUrn} allows the approximate numerical evaluation of the probability mass function of the Wallenius distribution. In a classical setting, this makes feasible the computation of the MLE. 
In a Bayesian setting this enables the implementation of standard MCMC algorithms, such as the Metropolis-Hastings sampler. 
Nonetheless, we deem more appropriate to use the ABC approach illustrated in this section for several reasons. 
First, the output of the Bayesian approach is far richer than the one available in a classical setting. For instance, in Section \ref{subsec:journals} we are able to easily compute important summaries of the posterior distribution, i.e. the probability $p_{ij}=\Pr(\omega_i>\omega_j)$. 
Second, standard MCMC methods require repeated evaluations of the likelihood function. This could lead to an unsustainable computational burden compared to ABC. 
Last but not least, we have performed a simulation study regarding the behaviour of the maximum likelihood estimator of the vector 
$\bm{\omega}$ and we noticed that it typically tends to produce unreliable and unstable estimates when the ``true'' $\bm{\omega}$ is close to the boundary of the simplex and/or when the number of categories is large.

%\subsubsection*{Identifiability}
 
% We conclude this section with a cautionary remark about parameter identifiability. First of all, we recall that the quantity $d$ in the  p.m.f. of the  Wallenius distribution (defined in equation \eqref{wal-gen}) depends on the priority parameters $\boldsymbol{\omega}$. In particular
%$$d= \sum_{j=1}^c \omega_j (m_j - x_j).$$
%When we consider two different vectors $\boldsymbol{\omega}\prime$ and $\boldsymbol{\omega}$ such that $\boldsymbol{\omega}\prime = \kappa \boldsymbol{\omega}$ for $\kappa > 0$, we have that 
%\begin{equation}\label{ident}
%\frac{\omega_j'}{d'}=\frac{\kappa\omega}{\sum_{j=1}^c \kappa \omega_j (m_j - x_j)}=\frac{\omega}{\sum_{j=1}^c \omega_j (m_j - x_j)}=\frac{\omega_j}{d}
%\end{equation}
%where $d'$ and $d$ are computed respectively with $\boldsymbol{\omega}\prime$ and $\boldsymbol{\omega}$. Equation \eqref{ident} implies that the p.m.f. of the Wallenius distribution doesn't change if we consider the vector of priorities $\boldsymbol{\omega}\prime$ instead of $\boldsymbol{\omega}$. This induces an identifiability issue which can be tackled by normalizing the vector $\boldsymbol{\omega}$ with respect to one component, constraining the estimation to the remaining $(c-1)$ components of $\boldsymbol{\omega}$. In the examples in the following sections, we normalized, in general, with respect to the last value $\omega_c$, the only exception being the third simulation study. 
 
%%%%%%%%%%%%%%%%%%%%%%%%%%%%%%%%%%%%%%%%%%%%%%%%%%%%%%
%% 	 	Simulation 		%%%
%%%%%%%%%%%%%%%%%%%%%%%%%%%%%%%%%%%%%%%%%%%%%%%%%%%%%%
\section{Simulation Study}
\label{Simu}

In order to test Algorithm §\ref{ABC-rej} with the summary statistics shown in Section \ref{BayesInf}, we have conducted an extensive simulation study, with different scenarios. We performed $20$ repeated simulations of $k$ draws from the Wallenius distribution where each draw consists of a number $n_h$ $(h=1, \cdots, k)$ of balls. We use the prior distribution defined in Section \ref{BayesInf}, i.e. a Dirichlet prior $\mathcal{D}ir(1,\dots,1)$. As already stated in Section \ref{BayesInf}, we use the summary statistics and the distance in variation defined in equations \eqref{sum-stat} and \eqref{eq:dist}. The tolerance level $\varepsilon$ has been chosen with a pilot simulation where $10^5$ values have been simulated by fixing the tolerance level to a very large value. Then, the distribution of the distances from the true values has been studied. The tolerance level is fixed as a small quantile of this distribution (it is common practice to fix it as the quantile of level $0.05$). The complete procedure will be described in the following.
%We have then performed a pilot simulation in order to set the tolerance level $\varepsilon$: we have proposed values and studied the (approximated) distribution of the relative threshold and then picked up the $\varepsilon$ corresponding to the $5$-th lower quantile. 
%We have applied Algorithm \ref{ABC-rej} until we have obtained $10^4$ accepted values. 
%
The simulated experiments have been performed for different values of $c$, ranging between $2$ and $20$, and using three configurations for both $\textbf{m}$ and $\boldsymbol{\omega}$, as explained below: 
\begin{itemize}
\item same number of balls for each colour, i.e. ${m_j=m}$, $j=1, \dots,c$;  uniform importance weights, i.e. ${\omega_j=\omega}$, $j=1, \dots,c$;  
\item increasing values for $m_j$'s (all the integers between $1$ and $c$) and ${\omega}$'s (all the integers between $1$ and $c$, normalized to sum to one), $j=1, \dots, c$;
\item increasing values for $m_j$'s (all the integers between $1$ and $c$) and decreasing values for the ${\omega}$'s (all the integers between $c$ and $1$, normalized to sum to one), $j=1, \dots, c$;
\end{itemize}
Finally, we have used three different sample sizes, namely $k=5$, $k=50$ and $k=1000$. 
The value of $n_h$'s has been taken to be half the total number of balls in the urn. The results are available in Tables \ref{tab:omega1}, \ref{tab:omega2} and \ref{tab:omega3}. 

Surprisingly, as the sample size $k$ increases, the root mean squared error ($RMSE$) remains relatively stable. 
Results are less accurate for those configurations where both $\boldsymbol{\omega}$ and $\textbf{m}$ are uniform, while they are more accurate for configurations where $\boldsymbol{\omega}$ and $\textbf{m}$ follow an opposite ordering. This may be explained by observing that data are carrying more information on $\boldsymbol{\omega}$ in this situation. 

The $RMSE$ is decreasing almost everywhere as the value of $c$ increases: the only case where this is not true is the case of both $\boldsymbol{\omega}$ and $\textbf{m}$ uniform. 
This may suggest that the Wallenius distribution does not perform well when the ``true'' model is the simpler classical multivariate Hypergeometric model, especially when the number of categories $c$ is large.
%Forse dovremmo usare un'altra prior ....
Table \ref{tab:omega1}, \ref{tab:omega2} and \ref{tab:omega3} also show the average acceptance rates of the ABC algorithm used in the simulation experiments. The acceptance rate depends on the value of the tolerance level $\epsilon$ chosen in the experiment: we have followed the strategy described in \cite{allingham2009bayesian}, where a pilot run is done to study the distribution of the distance between the summary statistics computed on the observed data and on the simulated data. Then, $\varepsilon$ is chosen to be a quantile of the empirical distribution of this distance. We have chosen to consider the quantile of level $0.05$. With this automatic choice of $\varepsilon$ we obtain an acceptance rate of about $0.01-0.02$ on average. We obtained lower acceptance rates in the case of a small number of colours. These rates are compatible with the average tolerance level. It could be possible to reduce the $RMSE$ by reducing the tolerance level $\varepsilon$, however there is a balance between the goodness of the approximation and the computational cost. In an applied context, it is always advisable to compare several tolerance levels. We will propose this comparison in Section \ref{RealData}. In this context, we use only one threshold $\varepsilon$ (in the automatic way above described) to focus the analysis on a Monte Carlo comparison by varying the sample size and the number of colours in the urn.

As a conclusive remark of the section, we have performed a sensitivity analysis regarding the common hyperparameter of the Dirichlet prior. For values ranging from $1/c$ (the choice suggested in \cite{berger2015})  and $1$ (the uniform prior), we have always obtained similar results in terms of RMSE, showing a sort of robustness of the model, at least with respect to this particular aspect.
\begin{table}[H]
\centering
\caption{Simulation study; Three different sample sizes: $k=5$, $k=50$, $k=1000$. Twenty replications of the experiment with uniform true values for $\bm{\omega}$ and $\bm{m}$ for each size of categories ($c=2,\dots,20$). The root mean squared error and the average acceptance rate are reported.}
\label{tab:omega1}
\vspace{0.1in}
\begin{tabular}{c|cc|cc|cc}
\textbf{}   & \textbf{k=5}  & \textbf{}          & \textbf{k=50} & \textbf{}          & \textbf{k=1000} & \textbf{}          \\
$\bm{c}$    & \textit{RMSE} & \textit{acc. rate} & \textit{RMSE} & \textit{acc. rate} & \textit{RMSE}   & \textit{acc. rate} \\ \hline
\textbf{2}  & 0.7084        & 0.0018             & 0.7071        & 0.0017             & 0.7071          & 0.0016             \\
\textbf{3}  & 0.2922        & 0.0057             & 0.2887        & 0.0057             & 0.2886          & 0.0057             \\
\textbf{4}  & 0.1714        & 0.0082             & 0.1667        & 0.0080             & 0.1667          & 0.0080             \\
\textbf{5}  & 0.1118        & 0.0096             & 0.1119        & 0.0095             & 0.1119          & 0.0094             \\
\textbf{6}  & 0.0912        & 0.0104             & 0.0819        & 0.0102             & 0.0818          & 0.0102             \\
\textbf{7}  & 0.0811        & 0.0108             & 0.0634        & 0.0110             & 0.0632          & 0.0109             \\
\textbf{8}  & 0.0662        & 0.0115             & 0.0511        & 0.0113             & 0.0508          & 0.0114             \\
\textbf{9}  & 0.0576        & 0.0119             & 0.0423        & 0.0117             & 0.0420          & 0.0117             \\
\textbf{10} & 0.0534        & 0.0121             & 0.0356        & 0.0121             & 0.0357          & 0.0121             \\
\textbf{15} & 0.1326        & 0.0132             & 0.1292        & 0.0131             & 0.1292          & 0.0131             \\
\textbf{20} & 0.1845        & 0.0138             & 0.1830        & 0.0136             & 0.1829          & 0.0136            
\end{tabular}
\end{table}

\begin{table}[H]
\centering
\caption{Simulation study; Three different sample sizes: $k=5$, $k=50$, $k=1000$. Twenty replications of the experiment with increasing values for  $\bm{\omega}$ and $\bm{m}$ for each size of categories ($c=2,\dots,20$). The root mean squared error and the average acceptance rate are reported.}\label{tab:omega2}
\begin{tabular}{c|cc|cc|cc}
\textbf{} & \textbf{K=5}  & \textbf{}          & \textbf{K=50} & \textbf{}          & \textbf{K=1000} & \textbf{}          \\
$\bm{c}$    & \textit{RMSE} & \textit{acc. rate} & \textit{RMSE} & \textit{acc. rate} & \textit{RMSE}   & \textit{acc. rate} \\ \hline
\textbf{2}  & 0.4792        & 0.0014             & 0.4590        & 0.0014             & 0.4702          & 0.0018             \\
\textbf{3}  & 0.4471        & 0.0048             & 0.6627        & 0.0067             & 0.6731          & 0.0070             \\
\textbf{4}  & 0.4547        & 0.0093             & 0.5150        & 0.0105             & 0.5176          & 0.0108             \\
\textbf{5}  & 0.4102        & 0.0115             & 0.4339        & 0.0119             & 0.4350          & 0.0120             \\
\textbf{6}  & 0.3461        & 0.0112             & 0.3866        & 0.0130             & 0.3902          & 0.0132             \\
\textbf{7}  & 0.3472        & 0.0124             & 0.3538        & 0.0143             & 0.3585          & 0.0144             \\
\textbf{8}  & 0.3061        & 0.0137             & 0.3255        & 0.0148             & 0.3238          & 0.0152             \\
\textbf{9}  & 0.2734        & 0.0144             & 0.2982        & 0.0153             & 0.3013          & 0.0153             \\
\textbf{10} & 0.2590        & 0.0172             & 0.2806        & 0.0158             & 0.2816          & 0.0159             \\
\textbf{15} & 0.1971        & 0.0189             & 0.2153        & 0.0170             & 0.2172          & 0.0171             \\
\textbf{20} & 0.1628        & 0.0198             & 0.1803        & 0.0177             & 0.1628          & 0.0177            
\end{tabular}
\end{table}

\begin{table}[H]
\centering
\caption{Simulation study; Three different sample sizes: $k=5$, $k=50$, $k=1000$. Twenty replications of the experiment with increasing true values for $\bm{m}$ and decreasing values for $\bm{\omega}$ for each size of categories ($c=2,\dots,20$). The root mean squared error and the average acceptance rate are reported.}

\label{tab:omega3}
\begin{tabular}{c|cc|cc|cc}
\textbf{}   & \textbf{K=5}  & \textbf{}          & \textbf{K=50} & \textbf{}          & \textbf{K=1000} & \textbf{}          \\
$\bm{c}$   & \textit{RMSE} & \textit{acc. rate} & \textit{RMSE} & \textit{acc. rate} & \textit{RMSE}   & \textit{acc. rate} \\ \hline
\textbf{2}  & 0.0117        & 0.0014             & 0.0013        & 0.0013             & 0.0013          & 0.0017             \\
\textbf{3}  & 0.1464        & 0.0052             & 0.2428        & 0.0070             & 0.2502          & 0.0071             \\
\textbf{4}  & 0.0888        & 0.0092             & 0.0975        & 0.0107             & 0.0982          & 0.0109             \\
\textbf{5}  & 0.0633        & 0.0116             & 0.0579        & 0.0120             & 0.0586          & 0.0120             \\
\textbf{6}  & 0.0890        & 0.0128             & 0.0741        & 0.0132             & 0.0738          & 0.0132             \\
\textbf{7}  & 0.0882        & 0.0138             & 0.0724        & 0.0143             & 0.0752          & 0.0146             \\
\textbf{8}  & 0.0961        & 0.0144             & 0.0693        & 0.0152             & 0.0690          & 0.0152             \\
\textbf{9}  & 0.0907        & 0.0148             & 0.0715        & 0.0152             & 0.0695          & 0.0154             \\
\textbf{10} & 0.0875        & 0.0154             & 0.0709        & 0.0157             & 0.0725          & 0.0158             \\
\textbf{15} & 0.0940        & 0.0172             & 0.0753        & 0.0171             & 0.0748          & 0.0173             \\
\textbf{20} & 0.0891        & 0.0182             & 0.0732        & 0.0179             & 0.0731          & 0.0177            
\end{tabular}
\end{table}

%%%%%%%%%%%%%%%%%%%%%%%%%%%%%%%%%%%%%%%%%%%%%%%%%%%%%%
%% 	 	Real Data 		%%%
%%%%%%%%%%%%%%%%%%%%%%%%%%%%%%%%%%%%%%%%%%%%%%%%%%%%%%
\section{Real Data Applications}
\label{RealData}

We now apply the proposed approach to two real datasets, in order to assess the applicability and the performance of the algorithm. 
In both cases, we obtain the ratings of a group of individuals about specific elements from a list. Each individual may choose the number of elements to rate. The elements are then grouped in categories and the goal is to provide a ranking of the categories. By using the urn terminology of Section \ref{Prelim}, the categories are the colours and each element from the list is a ball; the aim of the analysis is to perform inference on the importance weights of each colour.  

\subsection{Movies dataset}
\label{subsec:movies}

\begin{figure}
\includegraphics{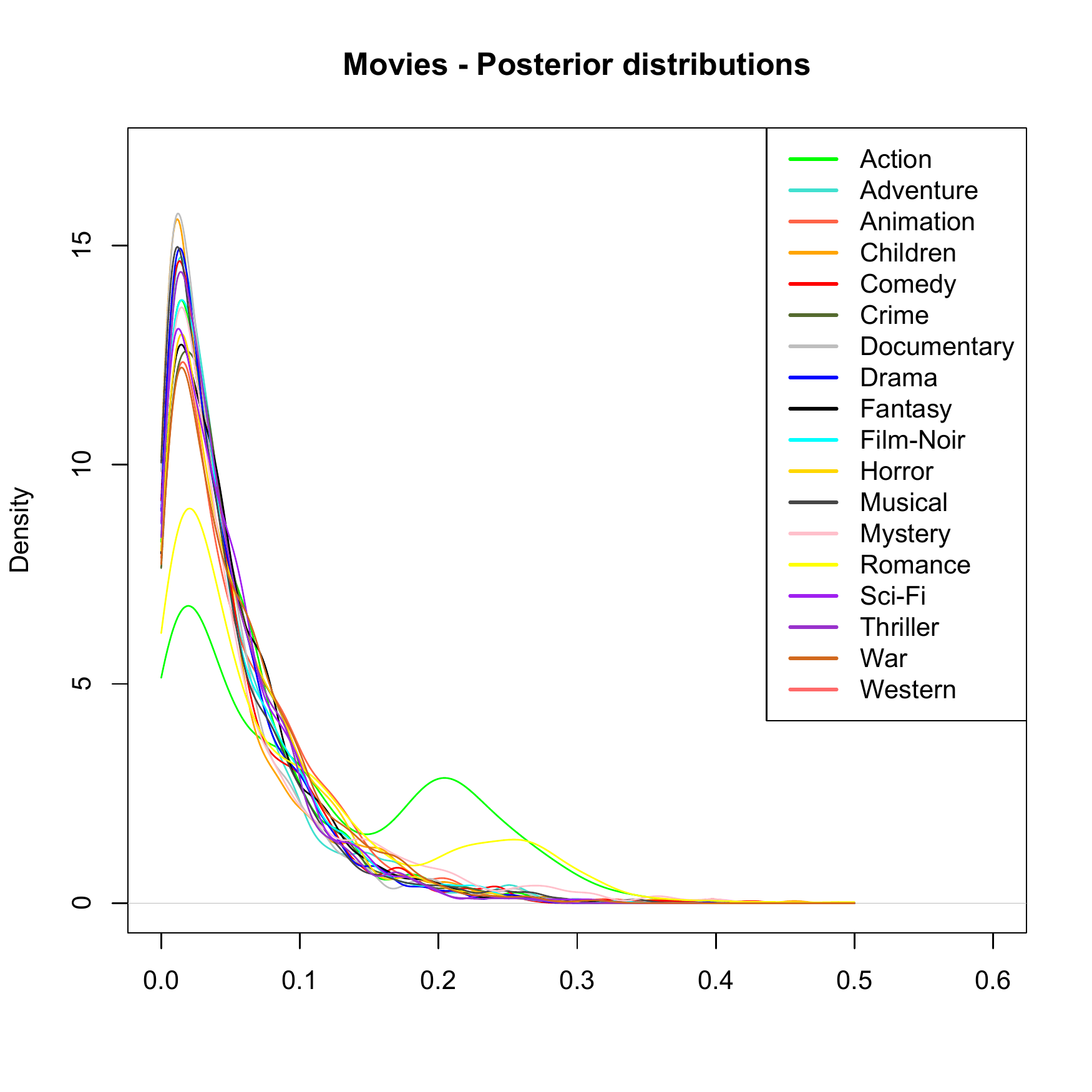}
\caption{Approximations of the posterior distributions of the weights $\boldsymbol{\omega}$ for each category included in the Movies dataset.}
\label{fig:movies}
\end{figure}

This dataset describes 5-star (with half-star increments) rating from MovieLens, a movie recommendation service (http://grouplens.org/datasets/movielens/). The dataset may change over time. We consider the dataset which contains 105,339 ratings across 10,329 movies. These data were created by 668 users between April 03, 1996 and January 09, 2016. This dataset was generated on January 11, 2016. Users were randomly selected by MovieLens, with no demographic information, and each of them has rated at least 20 movies. The movies in the dataset were described by genre, following the \textit{IMDb} information (https://www.themoviedb.org/); nineteen genres were considered in the dataset, including a ``no genre'' category; we have decided to eliminate the empty category from the analysis. In this case, we consider a movie to be "good" if its rating is at least $3.5$ stars. Therefore, the vector $X_n$ represents the frequencies of "good movies" in each category. Each film may be described by more than one genre. In this case we have proceeded as follows: we have ordered the genres in terms of their generality and then assigned to the movie the least general genre with which it was described. We have decided the following order (from the less general to the most general): Animation $\rightarrow$ Children $\rightarrow$ Musical $\rightarrow$ Documentary $\rightarrow$ Horror $\rightarrow$ Sci-Fi $\rightarrow$ Film Noir $\rightarrow$ Crime $\rightarrow$ Fantasy $\rightarrow$ War $\rightarrow$ Western $\rightarrow$ Mistery $\rightarrow$ Action $\rightarrow$ Thriller $\rightarrow$ Adventure $\rightarrow$ Romance $\rightarrow$ Comedy $\rightarrow$ Drama. Of course, this is an experimental choice, which may affect the results. Since the movies can be cross-classified, an interesting (and more realistic) development would be considering a model which can take into account this feature; this is left for further research. We have then replicated the same prior choice and the same choices of distance and vector of summary statistics described in Section \ref{Simu}. The tolerance level $\varepsilon$ has been chosen with a pilot simulation in order to produce a sample of size $10^5$, as described in Section \ref{Simu}. In this particular case, we have used $\varepsilon=0.5$. Table \ref{tab:movies} displays the posterior mean estimates of the vector of importance weights $\boldsymbol{\omega}$. The importance weights seem to be very close, with small differences among them. This suggests that there is not a category which is particularly popular. Nonetheless, we can observe a slightly preference for the \textit{Action} and \textit{Sci-Fi} genres and less interest in the \textit{Fantasy}, \textit{War} and \textit{Drama} genres. We believe that this similarity in the importance weights is due to an excessive number of categories in the movies dataset. In this setting the graphical comparison of the marginal posterior distributions can provide a better insight on the customer preferences. Figure \ref{fig:movies} shows that there is more variability in the users preferences to choose a particular movie genre, such as \textit{Action} or \textit{Romance}.

\begin{table}[h]
\centering
\caption{Posterior mean estimates and standard deviations (in brackets) of the vector of importance weights $\boldsymbol{\omega}$ for each genre with tolerance level $\varepsilon=0.5$. }
\label{tab:movies}
\vspace{0.1in}
\begin{tabular}{c|c|c|c}
                     & $\boldsymbol{\omega}$ &                    & $\boldsymbol{\omega}$ \\ \hline
\textbf{Action}      & 0.102                 & \textbf{Crime}     & 0.050                 \\
\textit{}            & \textit{(0.090)}      & \textit{}          & \textit{(0.055)}      \\
\textbf{Sci-Fi}      & 0.086                 & \textbf{Thriller}  & 0.050                 \\
\textit{}            & \textit{(0.089)}      & \textit{}          & \textit{(0.047)}      \\
\textbf{Romance}     & 0.059                 & \textbf{Horror}    & 0.050                 \\
\textit{}            & \textit{(0.068)}      & \textit{}          & \textit{(0.049)}      \\
\textbf{Children}    & 0.056                 & \textbf{Animation} & 0.049                 \\
\textit{}            & \textit{(0.054)}      & \textit{}          & \textit{(0.051)}      \\
\textbf{Western}     & 0.055                 & \textbf{Comedy}    & 0.049                 \\
\textit{}            & \textit{(0.051)}      & \textit{}          & \textit{(0.055)}      \\
\textbf{Musical}     & 0.052                 & \textbf{Mystery}   & 0.048                 \\
\textit{}            & \textit{(0.048)}      & \textit{}          & \textit{(0.052)}      \\
\textbf{Documentary} & 0.051                 & \textbf{Fantasy}   & 0.047                 \\
\textit{}            & \textit{(0.048)}      & \textit{}          & \textit{(0.046)}      \\
\textbf{Film-Noir}   & 0.051                 & \textbf{War}       & 0.047                 \\
\textit{}            & \textit{(0.048)}      & \textit{}          & \textit{(0.044)}      \\
\textbf{Adventure}   & 0.050                 & \textbf{Drama}     & 0.047                 \\
\textit{}            & \textit{(0.048)}      & \textit{}          & \textit{(0.051)}     
\end{tabular}
\end{table}

\subsection{Statistical Journals dataset}
\label{subsec:journals}

The scientific areas (or ``settori scientifici disciplinari'', S.S.D.) are a characterization used in the academic Italian system to classify knowledge in higher education. 
The sectors are determined by the Italian Ministry of Education. In particular, there are 367 S.S.D., divided into 14 macro-areas and each member of the academic staff pertains to a single sector. We have performed a survey on the preferences of the researchers in Statistics (Sector SECS-S/01) of Italian universities about the available scientific journals. It should be noted that researchers in Probability and Mathematical Statistics, Medical, Economic and Social Statistics are not included in this survey, because they pertain to different sectors. We have considered only staff with both teaching and research contracts. Postdoctoral fellows and PhD students have been excluded. In this survey we have used the 2015 ``Statistics and Probability'' list of journals of the Institute for Scientific Information (ISI). We have asked to SESC-S/01 researchers to indicate their preferences in this list, between a minimum of ten and a maximum of twenty. One difference from the Movies example of Section \ref{subsec:movies} is that the participants do not have to indicate the level of their preference, only a list of journals which each of the participants considers either

\begin{itemize}
	\item prestigious and/or
	\item likely for a potential submission and/or
	\item professionally significant (in terms of frequency of readings).
\end{itemize}

The survey was conducted between 25th October 2016 and 4th November 2016. We have collected 174 responses, distributed, in terms of role, as follows: 49 Full professors (Professori Ordinari), 72 Associate Professors (Professori Associati) and 53 Assistant Professors, both fixed-term and tenure-track (Ricercatori a tempo indeterminato e a tempo determinato). We have then grouped the journals by category, considering five main classes of interest: \textit{Methodology, Probability, Applied Statistics, Computational Statistics} and \textit{Econometrics and Finance}. The list of journals and relative category is available in the Appendix. Among the 124 journals available in the ``Statistics and Probability'' ISI list, we have classified 23 journals in \textit{Probability}, 45 in \textit{Methodology}, 34 in \textit{Applied Statistics}, 9 in \textit{Computational Statistics} and 13 in \textit{Econometrics and Finance}. We assume the Wallenius distribution for modelling the dataset, where $c$ represents the number of the categories. The preferences of each respondent are summarized in a vector where the position of each entry represents the number of journals falling in the corresponding category. We consider that this vector is a realization of the Wallenius distribution. 

\begin{table}[]
\centering
\caption{Each entry $p_{ij}$ of the matrix represents the ABC approximation of $\Pr(\omega_i > \omega_j)$. The order is 1-Methodology, 2-Probability, 3-Applied Statistics, 4-Computational Statistics, 5-Econometrics and Finance.}
%Only the values of the upper right part of the matrix are shown because the corresponding values in the lower left side are $p_{ji}=1-p_{ij}$. }
\label{tab:journals_p}
\begin{tabular}{c|ccccc}
                    & \textbf{$\omega_1$} & \textbf{$\omega_2$} & \textbf{$\omega_3$} & \textbf{$\omega_4$} & \textbf{$\omega_5$} \\ \hline
\textbf{$\omega_1$} & -               & 1.000              & 0.999               & 0.394               & 1.000               \\
\textbf{$\omega_2$} &                     & - & 0.000               & 0.000               & 0.226               \\
\textbf{$\omega_3$} &                     &                     & -               & 0.104               & 0.951               \\
\textbf{$\omega_4$} &                     &                     &                     & -               & 0.992               \\
\textbf{$\omega_5$} &                     &                     &                     &                     & -              
\end{tabular}
\end{table}

\begin{figure}
\includegraphics[scale=0.9]{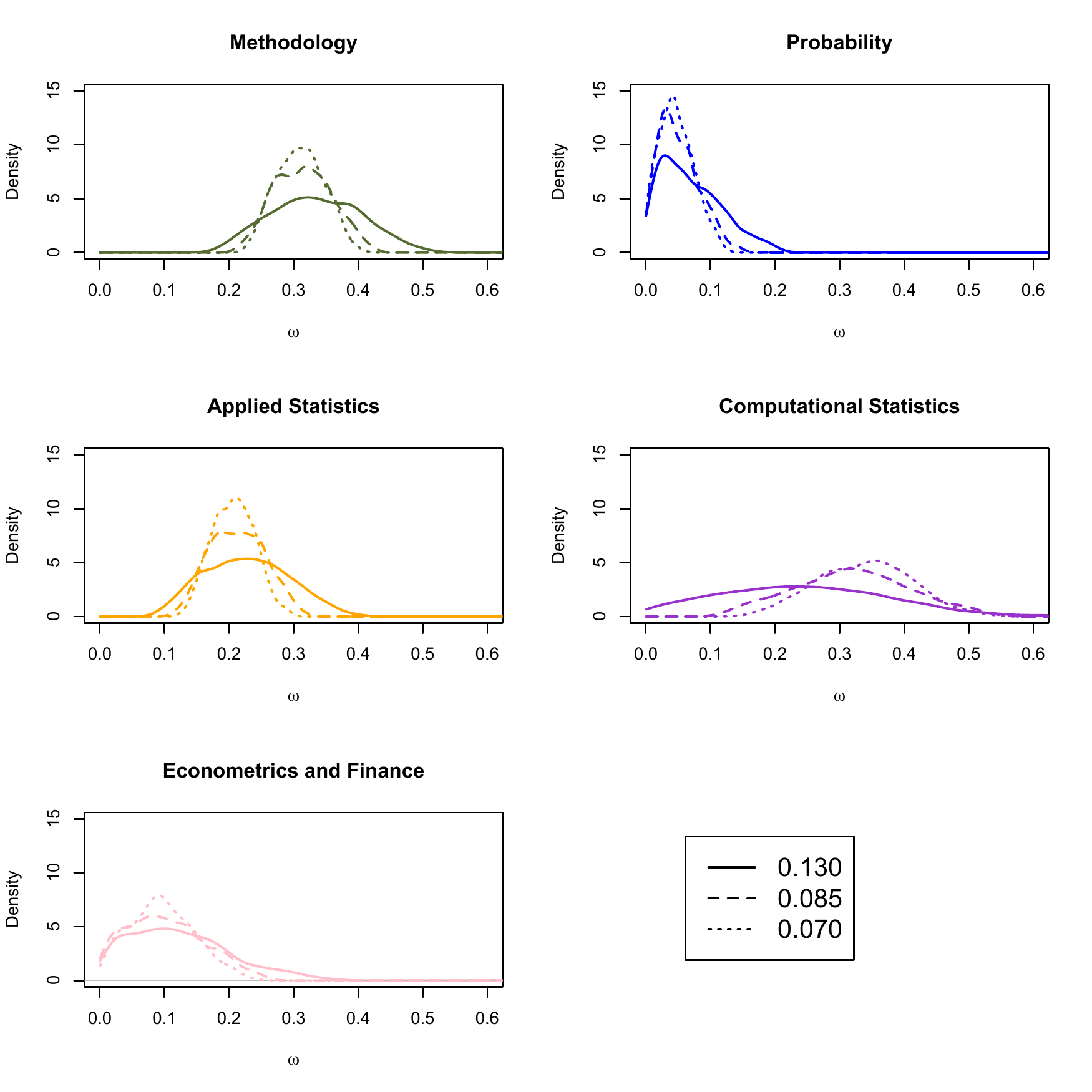}
\caption{Approximations of the posterior distributions of the weigths $\boldsymbol{\omega}$ for each category included in the Journals dataset.  Solid lines represent the approximations for $\varepsilon=0.130$, dashed lines for $\varepsilon=0.085$ and dotted lines for $\varepsilon=0.070$.}
\label{fig:journals}
\end{figure}

\begin{figure}
\centering
\includegraphics[scale=0.7]{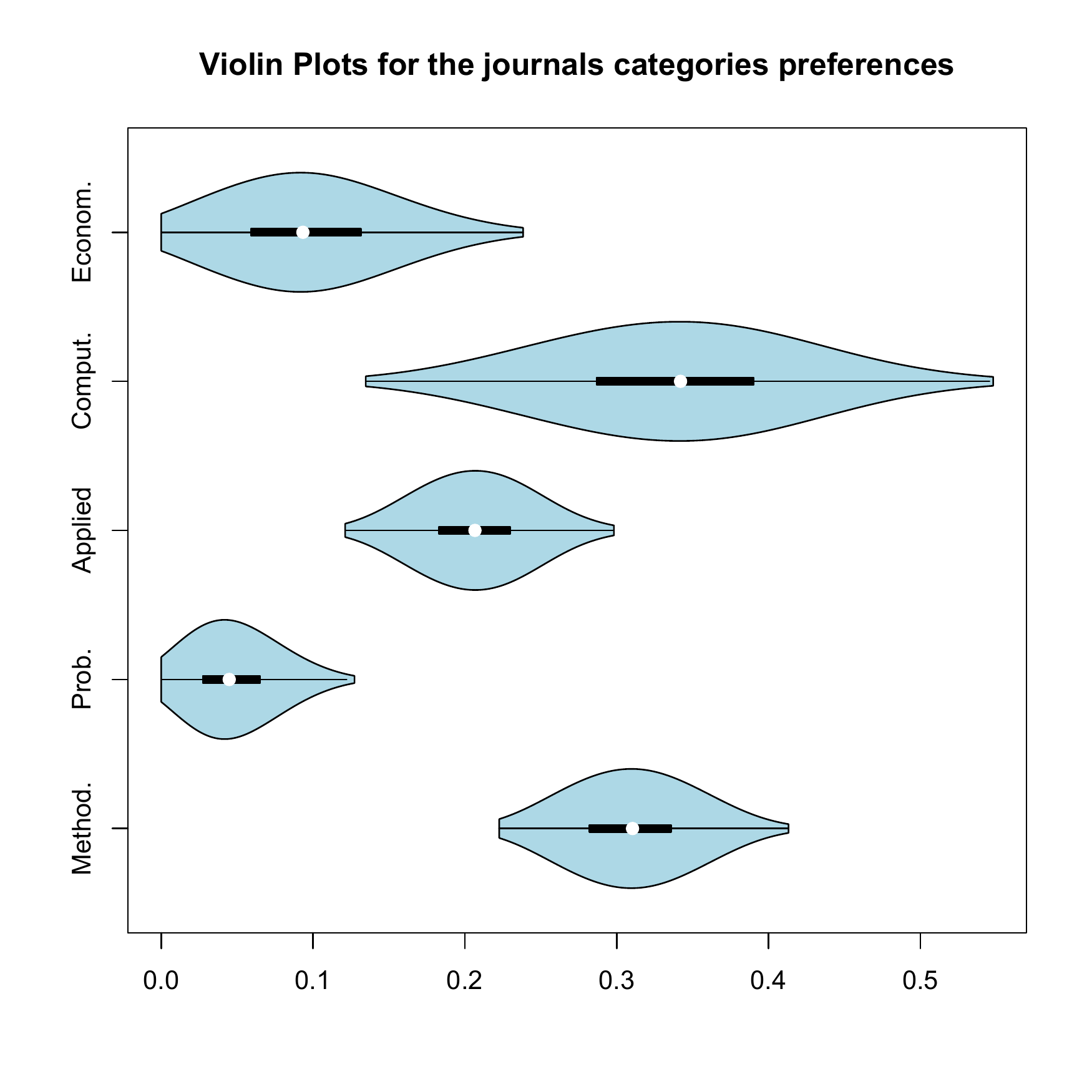}
\caption{Violin plots of the posterior distributions of the weigths $\boldsymbol{\omega}$ for each category included in the Journals dataset with $\varepsilon=0.070$. }
\label{fig:journals2}
\end{figure}

The results are available in Figure \ref{fig:journals}, Figure \ref{fig:journals2} and Table \ref{tab:journals}, which show that there seems to be a preference for the research in Methodological and Applied Statistics among the researchers in Statistics and less interest in journals of Probability. As already stated, this should highlight the fact that researchers in Mathematical Statistics and Probability do not pertain to the investigated sector. 
These results also show that the effect of a decrease of the tolerance level seems to be a concentration of the posterior distributions of the importance weights $\boldsymbol{\omega}$, except for the weight relative to the Computational journals, for which there is a shift.  As a possible explanation of this fact, one should consider that this category is under-represented in the list (at least, according our classification) with respect to the others. Table \ref{tab:journals_p} shows the estimated pair comparison probabilities for the journal categories.  

\begin{table}[H]
\centering
\caption{Posterior mean estimates and standard deviations (in brackets) of the vector of importance weights $\boldsymbol{\omega}$ for each category of journals and for different tolerance levels. }
\label{tab:journals}
\vspace{0.1in}
\begin{tabular}{r|c|c|c|c|c}
                                     & \textbf{Methodology} & \textbf{Probability} & \textbf{Applied} & \textbf{Computational} & \textbf{Econometrics} \\ \hline
\textbf{$\boldsymbol{\omega}$} & 0.335              & 0.070 & 0.228               & 0.244                      & 0.123                           \\
\textbf{$\varepsilon=0.130$}                            & \textit{(0.070)}  & \textit{(0.047)} &  \textit{(0.065)}  & \textit{(0.130)}          & \textit{(0.078)}               \\ \hline
\textbf{$\boldsymbol{\omega}$} & 0.315           & 0.051   & 0.213               & 0.320                      & 0.101                          \\
\textbf{$\varepsilon=0.085$}                            & \textit{(0.044)}   & \textit{(0.031)}  & \textit{(0.042)}   & \textit{(0.089)}          & \textit{(0.060)}                \\ \hline
\textbf{$\boldsymbol{\omega}$} & 0.310               &  0.048 & 0.207 & 0.339                      & 0.096                           \\
                               $\varepsilon=0.070$      & \textit{(0.037)}   & \textit{(0.027)}  & \textit{(0.033)}   & \textit{(0.073)}          & \textit{(0.050)}               
\end{tabular}
\end{table}

%%%%%%%%%%%%%%%%%%%%%%%%%%%%%%%%%%%%%%%%%%%%%%%%%%%%%%
%% 	 	Conclusion		%%%
%%%%%%%%%%%%%%%%%%%%%%%%%%%%%%%%%%%%%%%%%%%%%%%%%%%%%%
\section{Discussion}
\label{Concl}

In this paper we have considered the problem of ranking categories of items.  
We have proposed a novel model based on the Wallenius distribution. In terms of an urn scheme, it generalizes the Hypergeometric distribution with an additional vector of parameters $\boldsymbol{\omega}$, which represents the importance of the different types of balls in the urn. 

A referee noticed that ``the model assumes that the balls of the same colours (eg. the journals in the same category) are equally likely to be drawn.''  This assumption may not be justified, since, in the Journal example, journals in the same category may have different standing. This is exactly the reason why we propose the Wallenius model for ranking categories rather than single items; the weight $\boldsymbol{\omega}$ refers to the entire categories and they do not discriminate within categories. However, it is certainly of scientific interest to pursue the above issue and to conceive a nested model 
where items might be further ranked within categories; see, for example, \cite{inskip}.  In a Bayesian nonparametric setting, this approach could be further generalized by using nested non-exchangeable species sampling sequences, see \cite{Airoldi} and \cite{Bassetti}.  

So far the Wallenius model has been definitely under-employed, due to the analytical intractability of the probability mass function. In this work we proposed an approximate Bayesian computational algorithm which provides a fast and reliable approach to the estimation of the vector of priorities $\boldsymbol{\omega}$. Our method is easy to implement and it might be very useful in several statistical applications where balls are drawn from the urn in a biased fashion. Paradigmatic examples of the importance of the Wallenius model especially appear in auditing where transactions are randomly checked with probability proportional to their monetary value. We analysed two datasets concerning movies ratings and Italian academic statisticians' journal preferences. The ABC algorithm allows us to estimate the importance of movies categories or journal preferences under the assumption of a Wallenius generating model. Future work will focus on the use of the Wallenius distribution to other areas of application and on the estimation of the category multiplicities $\textbf{m}$ given the knowledge of the importance weights $\boldsymbol{\omega}$.

\section*{Acknowledgements}
The authors are very grateful to Martin Ridout for his valuable comments on a first draft of the paper. This project has been funded by the Royal Society International Exchanges Grant ``Empirical and Bootstrap Likelihood Procedures for Approximate Bayesian Inference''. F \textit{Applied Statistics}, 9 in \textit{Computational Statistics} and 13 in \textit{Econometrics and Finance}. We assume the Wallenius distribution for modelling the dataset, where $c$ represents the number of the categories. The preferences of each respondent are summarized in a vector where the position of each entry represents the number of journals falling in the corresponding category. We consider that this vector is a realization of the Wallenius distribution. 
abrizio Leisen was supported by the European Community's Seventh Framework Programme [FP7/2007-2013] under grant agreement no: 630677.

\vspace*{-8pt}

%\bibliographystyle{biometrika}
%\bibliographystyle{apalike}
%\bibliography{WalleniusBiblio}

%%%%%%%%%%%%%%%%%%%%%%%%%%%%%%%%%%%%%%%%%%%%%%%%%%%%%%
%%          Bibliography 		%%%
%%%%%%%%%%%%%%%%%%%%%%%%%%%%%%%%%%%%%%%%%%%%%%%%%%%%%%
%%%%%%%%%%%%%%%%%%%%%%%%%%%%%%%%%%%%%%%%%%%%%%%%%%%%%%
%%       	  Appendix			%%%
%%%%%%%%%%%%%%%%%%%%%%%%%%%%%%%%%%%%%%%%%%%%%%%%%%%%%%
\clearpage

\renewcommand{\thesection}{A}
\renewcommand{\theequation}{A.\arabic{equation}}
\renewcommand{\thefigure}{A.\arabic{figure}}
\renewcommand{\thetable}{A.\arabic{table}}
\setcounter{table}{0}
\setcounter{figure}{0}
\setcounter{equation}{0}

\section{Appendix}
\label{Proof}

\begin{table}[h]
\centering
\caption{Journals in the Probability category}
\vspace{0.1in}
\label{tab:probability}
\begin{tabular}{|l|}
\hline
\multicolumn{1}{|c|}{\textbf{Probability}}                                 \\ \hline
ADVANCES IN APPLIED PROBABILITY                                            \\ \hline
ANNALES DE L INSTITUT HENRI POINCARE - \\ PROBABILITES ET STATISTIQUES          \\ \hline
ANNALS OF APPLIED PROBABILITY                                              \\ \hline
ANNALS OF PROBABILITY                                                      \\ \hline
COMBINATORICS PROBABILITY and COMPUTING                                     \\ \hline
ELECTRONIC COMMUNICATIONS IN PROBABILITY                                   \\ \hline
ELECTRONIC JOURNAL OF PROBABILITY                                          \\ \hline
INFINITE DIMENSIONAL ANALYSIS QUANTUM PROBABILITY \\ AND RELATED TOPICS       \\ \hline
JOURNAL OF APPLIED PROBABILITY                                             \\ \hline
JOURNAL OF THEORETICAL PROBABILITY                                         \\ \hline
MARKOV PROCESSES AND RELATED FIELDS                                        \\ \hline
METHODOLOGY AND COMPUTING IN APPLIED PROBABILITY                           \\ \hline
PROBABILITY AND MATHEMATICAL STATISTICS-POLAND                             \\ \hline
PROBABILITY IN THE ENGINEERING AND \\ INFORMATIONAL SCIENCES                  \\ \hline
PROBABILITY THEORY AND RELATED FIELDS                                      \\ \hline
RANDOM MATRICES-THEORY AND APPLICATIONS                                    \\ \hline
STOCHASTIC ANALYSIS AND APPLICATIONS                                       \\ \hline
STOCHASTIC MODELS                                                          \\ \hline
STOCHASTIC PROCESSES AND THEIR APPLICATIONS                                \\ \hline
STOCHASTICS AND DYNAMICS                                                   \\ \hline
STOCHASTICS-AN INTERNATIONAL JOURNAL OF PROBABILITY \\ AND STOCHASTIC REPORTS \\ \hline
THEORY OF PROBABILITY AND ITS APPLICATIONS                                 \\ \hline
UTILITAS MATHEMATICA                                                       \\ \hline
\end{tabular}
\end{table}

\begin{table}[]
\centering
\caption{Journals in the Methodology category}
\label{tab:methodolody}
\footnotesize
\begin{tabular}{|l|}
\hline
\multicolumn{1}{|c|}{\textbf{Methodology}}                                           \\ \hline
ADVANCES IN DATA ANALYSIS AND CLASSIFICATION                              \\ \hline
ALEA-LATIN AMERICAN JOURNAL OF PROBABILITY AND \\  MATHEMATICAL STATISTICS    \\ \hline
AMERICAN STATISTICIAN                                                     \\ \hline
ANNALS OF STATISTICS                                                      \\ \hline
ANNALS OF THE INSTITUTE OF STATISTICAL MATHEMATICS                        \\ \hline
ANNUAL REVIEW OF STATISTICS AND ITS APPLICATION                           \\ \hline
ASTA-ADVANCES IN STATISTICAL ANALYSIS                                     \\ \hline
AUSTRALIAN and NEW ZEALAND JOURNAL OF STATISTICS                           \\ \hline
BAYESIAN ANALYSIS                                                         \\ \hline
BERNOULLI                                                                 \\ \hline
BIOMETRIKA                                                                \\ \hline
BRAZILIAN JOURNAL OF PROBABILITY AND STATISTICS                           \\ \hline
CANADIAN JOURNAL OF STATISTICS-REVUE CANADIENNE DE STATISTIQUE            \\ \hline
COMMUNICATIONS IN STATISTICS-THEORY AND METHODS                           \\ \hline
ELECTRONIC JOURNAL OF STATISTICS                                          \\ \hline
ESAIM-PROBABILITY AND STATISTICS                                          \\ \hline
EXTREMES                                                                  \\ \hline
FUZZY SETS AND SYSTEMS                                                    \\ \hline
HACETTEPE JOURNAL OF MATHEMATICS AND STATISTICS                           \\ \hline
INTERNATIONAL JOURNAL OF GAME THEORY                                      \\ \hline
INTERNATIONAL STATISTICAL REVIEW                                          \\ \hline
JOURNAL OF MULTIVARIATE ANALYSIS                                          \\ \hline
JOURNAL OF NONPARAMETRIC STATISTICS                                       \\ \hline
JOURNAL OF STATISTICAL PLANNING AND INFERENCE                             \\ \hline
JOURNAL OF THE AMERICAN STATISTICAL ASSOCIATION                           \\ \hline
JOURNAL OF THE KOREAN STATISTICAL SOCIETY                                 \\ \hline
JOURNAL OF THE ROYAL STATISTICAL SOCIETY SERIES B \\ STATISTICAL METHODOLOGY \\ \hline
JOURNAL OF TIME SERIES ANALYSIS                                           \\ \hline
LIFETIME DATA ANALYSIS                                                    \\ \hline
METRIKA                                                                   \\ \hline
REVSTAT-STATISTICAL JOURNAL                                               \\ \hline
SCANDINAVIAN JOURNAL OF STATISTICS                                        \\ \hline
SEQUENTIAL ANALYSIS-DESIGN METHODS AND APPLICATIONS                       \\ \hline
SPATIAL STATISTICS                                                        \\ \hline
STATISTICA NEERLANDICA                                                    \\ \hline
STATISTICA SINICA                                                         \\ \hline
STATISTICAL ANALYSIS AND DATA MINING                                      \\ \hline
STATISTICAL METHODOLOGY                                                   \\ \hline
STATISTICAL METHODS AND APPLICATIONS                                      \\ \hline
STATISTICAL MODELLING                                                     \\ \hline
STATISTICAL PAPERS                                                        \\ \hline
STATISTICAL SCIENCE                                                       \\ \hline
STATISTICS                                                                \\ \hline
STATISTICS and PROBABILITY LETTERS                                         \\ \hline
TEST                                                                      \\ \hline
\end{tabular}
\end{table}

\begin{table}[]
\centering
\caption{Journals in the Applied Statistics category}
\vspace{0.1in}
\label{tab:applied}
\begin{tabular}{|l|}
\hline
\multicolumn{1}{|c|}{\textbf{Applied Statistics}}                       \\ \hline
ANNALS OF APPLIED STATISTICS                                            \\ \hline
APPLIED STOCHASTIC MODELS IN BUSINESS AND INDUSTRY                      \\ \hline
BIOMETRICAL JOURNAL                                                     \\ \hline
BIOMETRICS                                                              \\ \hline
BIOSTATISTICS                                                           \\ \hline
BRITISH JOURNAL OF MATHEMATICAL and  STATISTICAL PSYCHOLOGY               \\ \hline
CHEMOMETRICS AND INTELLIGENT LABORATORY SYSTEMS                         \\ \hline
ENVIRONMENTAL AND ECOLOGICAL STATISTICS                                 \\ \hline
ENVIRONMETRICS                                                          \\ \hline
IEEE-ACM TRANSACTIONS ON COMPUTATIONAL BIOLOGY \\ AND BIONFORMATICS        \\ \hline
INTERNATIONAL JOURNAL OF BIOSTATISTICS                                  \\ \hline
JOURNAL OF AGRICULTURAL BIOLOGICAL \\ AND ENVIRONMENTAL STATISTICS         \\ \hline
JOURNAL OF APPLIED STATISTICS                                           \\ \hline
JOURNAL OF BIOPHARMACEUTICAL STATISTICS                                 \\ \hline
JOURNAL OF CHEMOMETRICS                                                 \\ \hline
JOURNAL OF COMPUTATIONAL BIOLOGY                                        \\ \hline
JOURNAL OF OFFICIAL STATISTICS                                          \\ \hline
JOURNAL OF QUALITY TECHNOLOGY                                           \\ \hline
JOURNAL OF THE ROYAL STATISTICAL SOCIETY SERIES A \\ STATISTICS IN SOCIETY \\ \hline
JOURNAL OF THE ROYAL STATISTICAL SOCIETY SERIES C \\ APPLIED STATISTICS    \\ \hline
MATHEMATICAL POPULATION STUDIES                                         \\ \hline
MULTIVARIATE BEHAVIORAL RESEARCH                                        \\ \hline
OPEN SYSTEMS and INFORMATION DYNAMICS                                    \\ \hline
PHARMACEUTICAL STATISTICS                                               \\ \hline
PROBABILISTIC ENGINEERING MECHANICS                                     \\ \hline
QUALITY ENGINEERING                                                     \\ \hline
SORT-STATISTICS AND OPERATIONS RESEARCH TRANSACTIONS                    \\ \hline
STATISTICAL APPLICATIONS IN GENETICS AND MOLECULAR BIOLOGY              \\ \hline
STATISTICAL METHODS IN MEDICAL RESEARCH                                 \\ \hline
STATISTICS IN BIOPHARMACEUTICAL RESEARCH                                \\ \hline
STATISTICS IN MEDICINE                                                  \\ \hline
STOCHASTIC ENVIRONMENTAL RESEARCH AND RISK ASSESSMENT                   \\ \hline
SURVEY METHODOLOGY                                                      \\ \hline
TECHNOMETRICS                                                           \\ \hline
\end{tabular}
\end{table}

\begin{table}[]
\centering
\caption{Journals in the Computational Statistics category}
\vspace{0.1in}
\label{tab:computational}
\begin{tabular}{|l|}
\hline
\multicolumn{1}{|c|}{\textbf{Computational Statistics}} \\ \hline
COMMUNICATIONS IN STATISTICS - \\ SIMULATION AND COMPUTATION \\ \hline
COMPUTATIONAL STATISTICS                                \\ \hline
COMPUTATIONAL STATISTICS and DATA ANALYSIS               \\ \hline
JOURNAL OF COMPUTATIONAL AND GRAPHICAL STATISTICS       \\ \hline
JOURNAL OF STATISTICAL COMPUTATION AND SIMULATION       \\ \hline
JOURNAL OF STATISTICAL SOFTWARE                         \\ \hline
R JOURNAL                                               \\ \hline
STATA JOURNAL                                           \\ \hline
STATISTICS AND COMPUTING                                \\ \hline
\end{tabular}
\end{table}

\begin{table}[]
\centering
\caption{Journal in the Econometrics and Financial Statistics category}
\vspace{0.1in}
\label{tab:econometrics}
\begin{tabular}{|l|}
\hline
\multicolumn{1}{|c|}{\textbf{Econometrics and Financial Statistics}} \\ \hline
ASTIN BULLETIN                                 \\ \hline
ECONOMETRIC REVIEWS                            \\ \hline
ECONOMETRIC THEORY                             \\ \hline
ECONOMETRICA                                   \\ \hline
ECONOMETRICS JOURNAL                           \\ \hline
FINANCE AND STOCHASTICS                        \\ \hline
INSURANCE MATHEMATICS and ECONOMICS             \\ \hline
JOURNAL OF BUSINESS and ECONOMIC STATISTICS     \\ \hline
LAW PROBABILITY and RISK                        \\ \hline
OXFORD BULLETIN OF ECONOMICS AND STATISTICS    \\ \hline
QUALITY and QUANTITY                            \\ \hline
QUALITY TECHNOLOGY AND QUANTITATIVE MANAGEMENT \\ \hline
SCANDINAVIAN ACTUARIAL JOURNAL                 \\ \hline
\end{tabular}
\end{table}

\end{document}